\documentclass{aa}  
\usepackage{graphicx}
\usepackage{txfonts}
\usepackage{color}
\usepackage[]{hyperref}
\hypersetup{
    colorlinks=true,
    linkcolor=blue,
    filecolor=blue,      
    urlcolor=blue,
    citecolor=blue
}
\usepackage{multicol}
\usepackage{color}
\usepackage{braket}
\usepackage{autobreak}
\usepackage{enumitem}
\usepackage{mathrsfs}
\usepackage{amsfonts}
\usepackage{amstext}
\usepackage{amsmath}
\usepackage{xcolor} 
\usepackage{natbib}
\usepackage{subfigure}
\def\fl#1 {\textcolor{cyan}{#1}\;}
\newcommand{\hi } {{\rm H}\,{\small\rm I}}

\begin{document} 

   \title{The baryonic mass-size relation of galaxies. I. A dichotomy in star-forming galaxy disks}
   \titlerunning{The baryonic mass-size relation of galaxies. I.}

   \author{Zichen Hua \inst{1, 2, 3},
          Federico Lelli \inst{1},
          Enrico Di Teodoro\inst{4, 1},
          Stacy McGaugh\inst{5},
          James Schombert\inst{6}
          }

   \institute{Arcetri Astrophysical Observatory, INAF, Largo Enrico Fermi 5, 50125, Florence, Italy
        \and
   Department of Astronomy, University of Science and Technology of China, Hefei 230026, China
        \and
            School of Astronomy and Space Sciences, University of Science and Technology of China, Hefei 230026, China
        \and
             Dipartimento di Fisica e Astronomia, Universit$\rm \grave{a}$ degli Studi di Firenze, 50019 Sesto Fiorentino, Italy
        \and
            Department of Astronomy, Case Western Reserve University, 10900 Euclid Avenue, Cleveland, OH 44106, USA
        \and
            Department of Physics, University of Oregon, Eugene, OR 97403, USA
        }
   \authorrunning{Z. Hua et al.}
   \date{Received XXX; accepted XXX}

  \abstract{
  The mass-size relations of galaxies are generally studied considering only stars or only gas separately. Here we study the baryonic mass-size relation of galaxies from the SPARC database, using the total baryonic mass ($M_{\rm bar}$) and the baryonic half-mass radius ($R_{\rm 50, bar}$). We find that SPARC galaxies define two distinct sequences in the $M_{\rm bar} - R_{\rm 50, bar}$ plane: one that formed by high-surface-density (HSD), star-dominated, Sa-to-Sc galaxies, and one by low-surface-density (LSD), gas-dominated, Sd-to-dI galaxies. The $M_{\rm bar} - R_{\rm 50, bar}$ relation of LSD galaxies has a slope close to 2, pointing to a constant average surface density, whereas that of HSD galaxies has a slope close to 1, indicating that less massive spirals are progressively more compact. Our results point to the existence of two types of star-forming galaxies that follow different evolutionary paths: HSD disks are very efficient in converting gas into stars, perhaps thanks to the efficient formation of non-axisymmetric structures (bars and spiral arms), whereas LSD disks are not. The HSD-LSD dichotomy is absent in the baryonic Tully-Fisher relation ($M_{\rm bar}$ versus flat circular velocity $V_{\rm f}$) but moderately seen in the angular-momentum relation (approximately $M_{\rm bar}$ versus $V_{\rm f}\times R_{\rm 50, bar}$), so it is driven by variations in $R_{\rm 50, bar}$ at fixed $M_{\rm bar}$. This fact suggests that the baryonic mass-size relation is the most effective empirical tool to distinguish different galaxy types and study their evolution.}
  

   \maketitle

\section{Introduction}\label{sec_intro}

One of the best studied scaling relations of galaxies is the stellar mass-size relation \citep[e.g.][]{Gadotti2009, Lange2015}, which relates the stellar mass ($M_\star$) to the effective radius ($R_{\rm 50, \star}$). The effective radius is generally defined as the radius that encompasses half of the stellar light or stellar mass of a galaxy, so it is also referred to as half-light or half-mass radii. Strictly speaking, the effective radius is not a measurement of `size' with the usual intuitive meaning of `maximum spatial extension' of an object because galaxies do not have a hard boundary. Rather, the half-mass radius is a measurement of how concentrated the stellar mass or stellar light is \citep{deVaucoulerus1948}. Other possible definitions of `galaxy sizes' consider isophotal or isodensity radii, which are taken at some fixed surface brightness (e.g. the Holmberg radius) or surface density value \citep[e.g.][]{Trujillo-R1}. For the sake of simplicity, hereafter we use the terms `effective radius', `half-mass radius', and `size' in an interchangeable way.

The stellar mass-size relation has been studied in connection with other galaxy properties, such as their morphology (e.g. \citealp{Shen2003-MR_SDSS, Bernardi2014, Schombert2006-structure}), surface brightness (e.g. \citealp{Greene2021-LSB-HSB}), specific star formation rates (e.g. \citealp{Nedkova2024-MR_evolution_decomp}), specific angular momentum \citep[e.g.][]{Kim2013-LSBG_spin, Rong2017-UDG}, interactions and mergers \citep[e.g.][]{Du2024-MR_origin, Liao2019-UDG}, and environment \citep[e.g.][]{Fernadez2013-MR_isolate, Rodriguez2021-MR_cluster, Afanasiev2023-MR_evolution_cluster}. In addition, one can probe the assembly history of galaxies by observing the evolution of the stellar mass-size relation with redshift \citep[e.g.][]{VandeW2014-MR_evolution, Roy2018-MR_evolution, Mowla2019-MR_evolution_Massive, Yang2021-MR_evolution, Afanasiev2023-MR_evolution_cluster, Nedkova2024-MR_evolution_decomp}. One can also study the distinct mass-size relations of different stellar components in galaxies, such as bulges and disks, in order to trace their evolutionary paths separately \citep[e.g.][]{Nedkova2024-MR_evolution_decomp}. The stellar mass-size relation has also been extensively investigated in the $\Lambda$ cold dark matter ($\Lambda$CDM) paradigm of galaxy formation, in particular in relation to the halo spin \citep{Mo1998-disk, Dutton2007-formation, Kim2013-LSBG_spin, Rong2017-UDG, Liao2019-UDG} and/or halo virial radius \citep{Kravtsov2013-MR_halo, Huang2017-MR_halo, Somerville2018-MR_halo, Rodriguez2021-MR_cluster}.

\begin{figure*}[th]
    \centering
    \includegraphics[width= 0.33\linewidth]{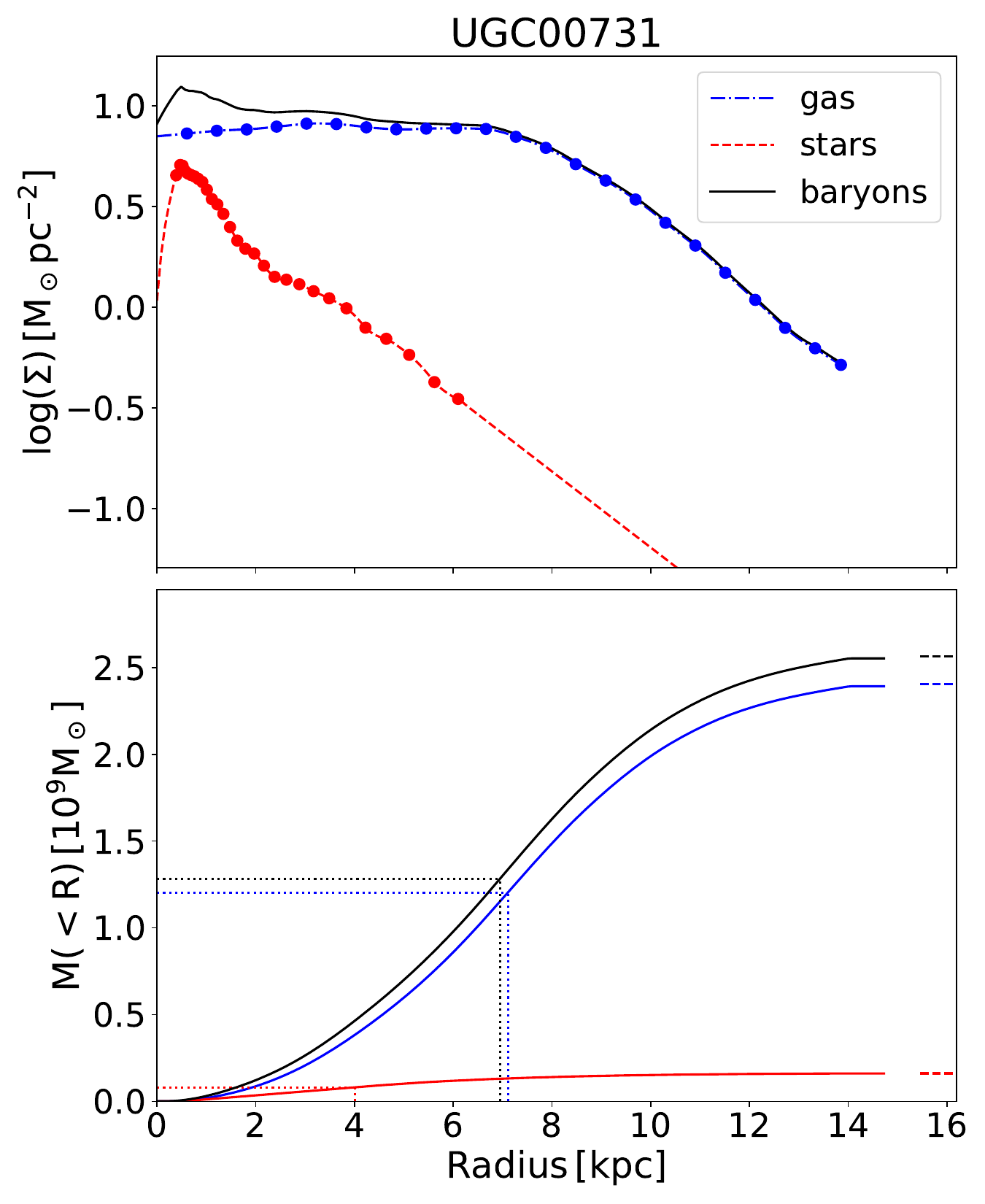}
    \includegraphics[width=0.33\linewidth]{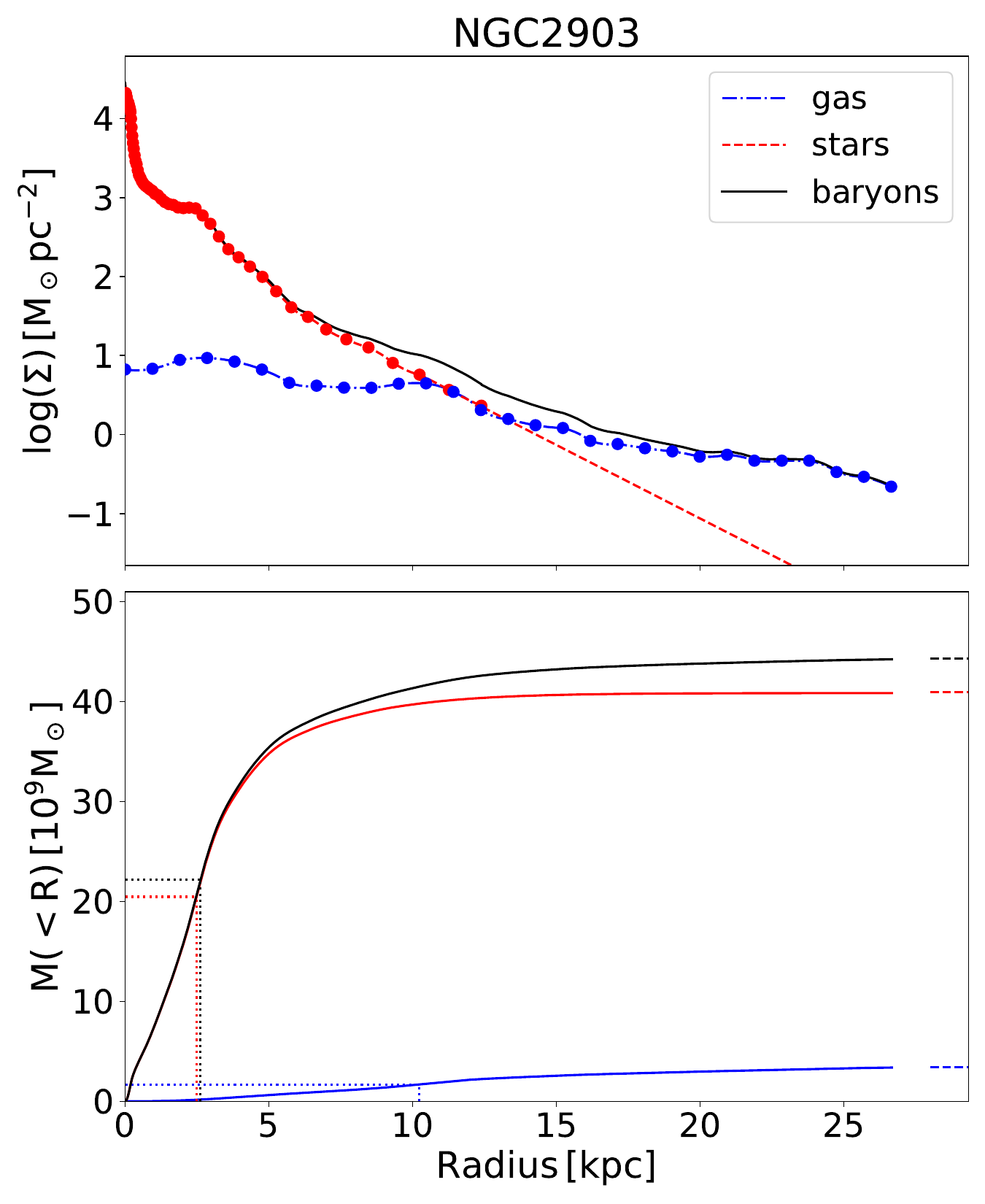}
    \includegraphics[width=0.33\linewidth]{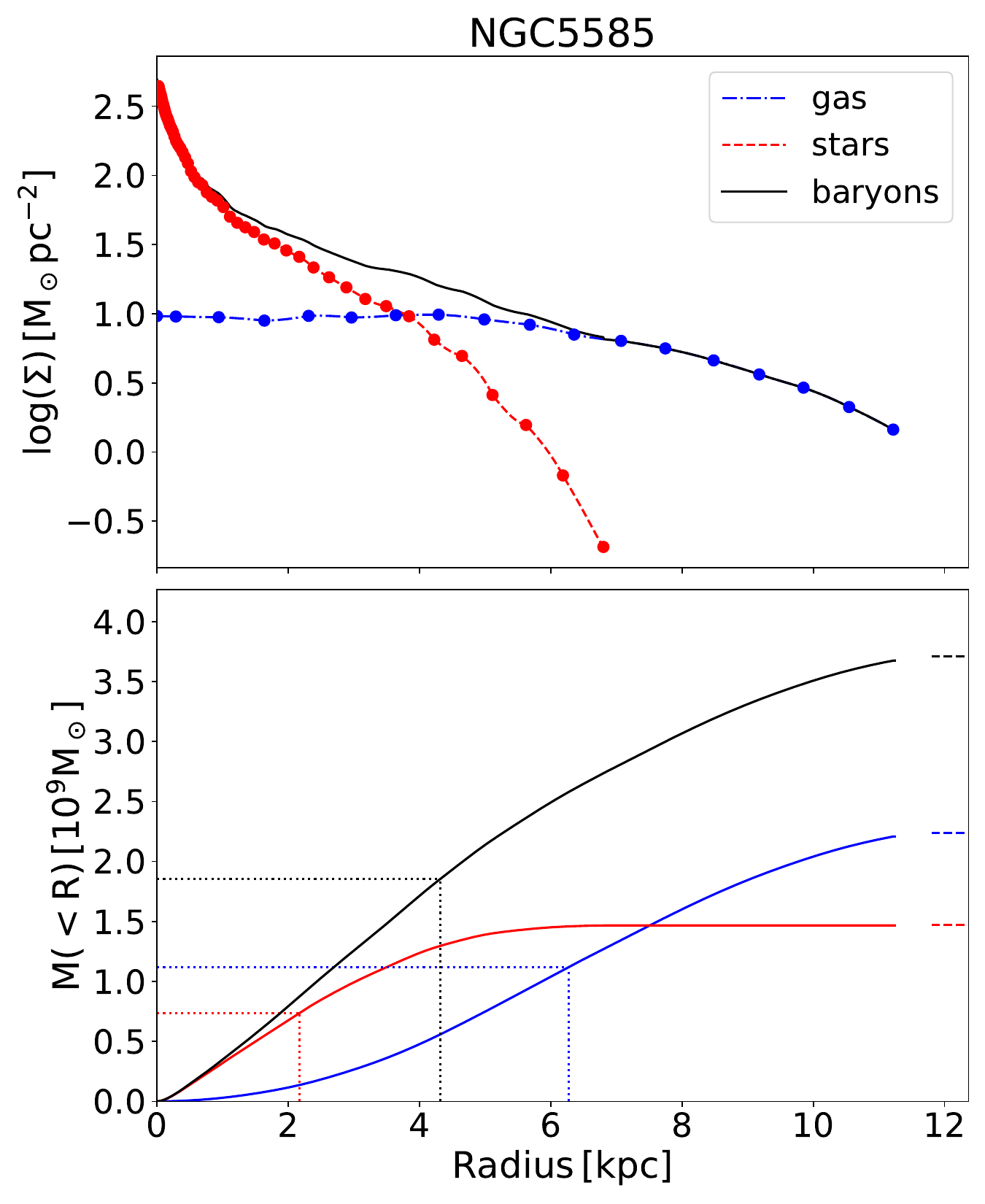}
    \caption{Measurements of $R_{\rm 50, gas}$, $R_{\rm 50, \star}$, and $R_{\rm 50, bar}$ for three example galaxies: a gas-dominated one (UGC\,731), a star-dominated one (NGC\,2903), and an intermediate case (NGC\,5585). The upper panels show the surface density profiles for gas (blue points), stars (red points), and total baryons (black line); the stellar profile was extrapolated at large radii with an exponential function (dashed red line). The lower panel shows the curve of growth of gas (blue line), stars (red line), and total baryons (black line); the dotted lines show the location of the corresponding half-mass radii.}
    \label{fig_derivation}
\end{figure*}

Another well-studied structural relation is the \hi\ mass-size relation of star-forming galaxies \citep{Broeils1997}. Given that atomic hydrogen largely dominates the gas budget of star-forming galaxies \citep[e.g.][]{Cortese2017}, the \hi\ mass-size relation is effectively the gaseous counterpart of the stellar mass-size relation. For historical reasons, the \hi\ radius ($R_{\rm HI}$) is not defined as the radius that contains 50$\%$ of the \hi\ mass, but rather as the radius where the \hi\ surface density equals $1 \ M_{\rm \odot}\,{\rm pc^{-2}}$ (after correction to face-on view), so it is effectively an isophotal radius. Interestingly, the \hi\ mass-size relation has a slope close to 2, indicating that the average surface densities of \hi\ disks are approximately constant \citep{Broeils1997, Verheijen2001, Swaters2002, Lelli2016-SPARC, Wang2016-M_HI-R_HI, Lutz2018-HIx, Gault2021-VLA_UDG}. This phenomenon may be related to the transformation of \hi\ into $\rm H_2$ (\citealp{Stevens2019_MHI-RHI_origin}), which is a crucial step in the star formation process.

Previous studies of the mass-size relation of galaxies focussed either only on stars or only on gas, but galaxies consist of both mass components. In particular, in star-forming dwarf galaxies, the gas mass can be comparable or even higher than the stellar mass \citep[e.g.][]{Lelli2022}, so neither the stellar mass-size relation nor the gaseous mass-size relation can thoroughly trace the matter distribution in their disks. In this paper, we study the baryonic mass-size relation of star-forming galaxies, linking the total baryonic mass ($M_{\rm bar}=M_\star + M_{\rm gas}$) with the baryonic effective radius ($R_{\rm 50, bar}$) that encloses half of $M_{\rm bar}$. 

The rest of this paper is structured as follows. In Section~\ref{sec_method}, we describe our sample and the derivation of $M_{\rm bar}$ and $R_{\rm 50, bar}$. In Section~\ref{sec_MR}, we present our results and find that star-forming galaxies follow two distinct sequences: one defined by high-surface-density (HSD) galaxies and one by low-surface-density (LSD) ones. In Section~\ref{sec_diss}, we discuss the HSD-LSD dichotomy in relation to the previous literature as well as to other galaxy scaling laws, such as the baryonic Tully-Fisher relation (BTFR) and the angular momentum relation (AMR). Finally, in Section~\ref{sec_conclusion}, we provide a brief summary of our results.

\begin{figure*}
    \centering
    \includegraphics[width=1\linewidth]{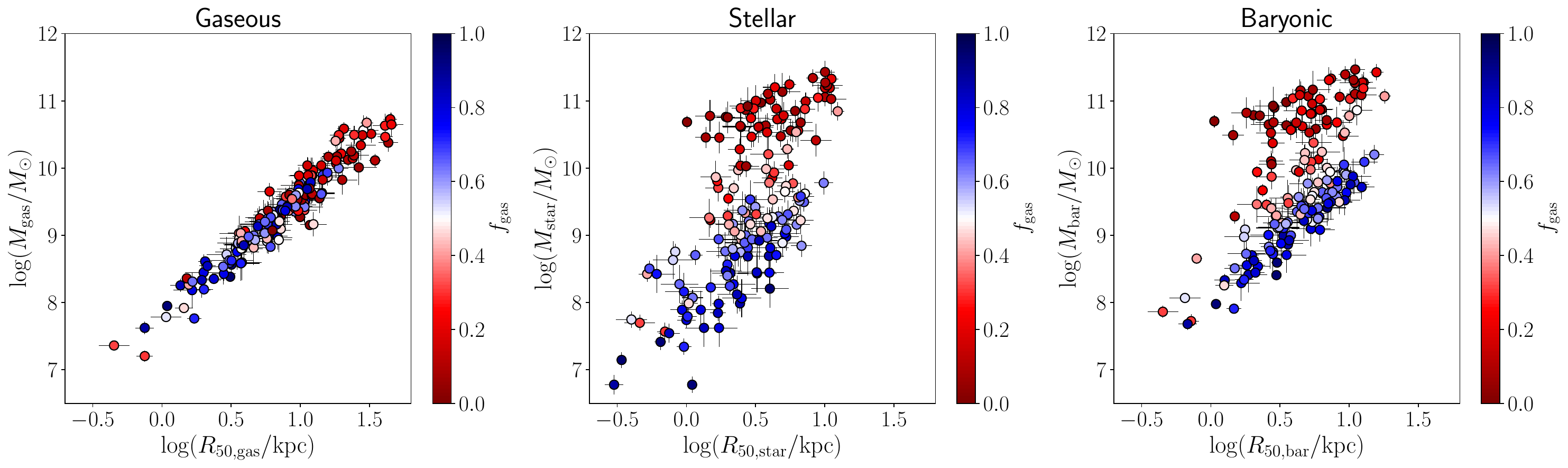}
    \caption{Gaseous (left panel), stellar (middle panel), and baryonic (right panel) mass-size relations of SPARC galaxies. Data points are colour-coded by the gas fraction $f_{\rm gas}$; the error bars denote the $1\sigma$ errors. The three panels span the same dynamic range on both axes. Two sequences are evident in the stellar and baryonic mass-size relations.}
    \label{fig_gas-star-bar}
\end{figure*}

\section{Data analysis}\label{sec_method}

\subsection{Gas and stellar surface density profiles}

The half-mass radius can be measured either by fitting a parametric function (such as the \citealt{Sersic1963} profile) to the matter distribution, or by constructing the so-called `curve of growth' (CoG) that gives the cumulative mass as a function of radius $R$. In general, the baryonic mass distribution of galaxies cannot be described by a simple parametric function, so we compute $R_{\rm 50, bar}$ by constructing the baryonic CoG (Fig.~\ref{fig_derivation}). This requires the gas and stellar surface density profiles corrected for face-on view. 

The Surface Photometry and Accurate Rotation Curves (SPARC) sample is the ideal dataset for our purposes because it contains 175 disk galaxies (Sa to dI)\footnote{The SPARC database contains three S0s (NGC\,4138, UGC\,2487, UGC\,6786) that are detected in the ultraviolet and/or in the H$\alpha$ line, suggesting recent star formation. For simplicity, in the following, we consider them as early-type spirals (Sa-Sc).} with both near-infrared (NIR) images at 3.6 $\rm \mu m$ from \textit{Spitzer}, tracing the stellar mass distribution, and interferometric \hi\ data, tracing the gas distribution and kinematics \citep{Lelli2016-SPARC}. The first SPARC data release \citep{Lelli2016-SPARC}, however, did not provide the \hi\ surface density profiles because the gravitational contribution of the gas disk (parametrized via the expected circular velocity $V_{\rm gas}$) was directly taken from previous works. Thus, we searched the literature and found \hi\ surface density profiles for 169 galaxies out of 175 galaxies. Unfortunately, for the remaining six galaxies (D512-2, D564-8, D631-7, NGC\,5907, NGC\,7339, UGC\,06818), the \hi\ rotation-curve references do not provide the \hi\ surface density profiles. The references for the \hi\ surface density profiles are listed in Table~\ref{tab_references}. In general, these authors use a consistent method to derive the \hi\ surface density profiles, that is taking azimuthal averages over a fixed set of rings, whose geometry (centre, position angle, and inclination angle $i$) is set by the \hi\ kinematics \citep[e.g.][]{Begeman1987}. An exception is represented by edge-on galaxies (with $i\gtrsim80^{\circ}$) for which the \hi\ surface density profiles are derived using the Lucy deconvolution method \citep[see][for details]{Swaters2002}. Our sample contains only 24 edge-on galaxies; they do not show any sign of systematic effect with respect to the rest of the sample.

To obtain the face-on \hi\ surface density profiles, we followed the same procedure applied by \citet{Lelli2016-SPARC} to the NIR surface brightness profiles. In short, we ran the task $\texttt{Rotmod}$ in the $\textsc{Gipsy}$ software \citep{Vogelaar2001}, which takes the observed surface brightness profile and total disk mass as inputs, then returns the expected circular velocity and face-on surface density as outputs. We used the same \hi\ masses and distances given in \citet{Lelli2016-SPARC}. \textcolor{black}{In running \texttt{Rotmod}}, we assumed an exponential vertical density profile with scale height of 100 pc \textcolor{black}{\citep[as in][]{Lelli2016-SPARC}}, but this assumption plays virtually no role \textcolor{black}{because we are interested in the face-on surface density integrated along the $z$ axis of the disk.} 

The \hi\ and NIR data have different spatial resolutions, so the resulting surface brightness profiles are sampled at different radii. To sum up the two components, the profiles are linearly interpolated and sampled on a common radial grid. Different choices in the interpolation play virtually no role in the final measurement of the half-mass radii. During this analysis, we also revisited some bulge-disk decompositions and outer extrapolations of the stellar profiles to ensure that the resulting CoGs are well-behaved with no unphysical jumps or discontinuities. We recall that SPARC bulge-disk decompositions are non-parametric and assign non-axisymmetric structures (bars, lenses, pseudo-bulges) to the stellar disk \citep{Lelli2016-SPARC}. \textcolor{black}{As we show in Sect.\,\ref{sec_bulges}, including or excluding bulges gives no differences at all in the baryonic mass-size relation of LSD galaxies (mostly Sd-to-dI types) and only small differences in the one of HSD galaxies (mostly Sa-to-Sc types), so it is clear that bulge-disk decompositions represent a second-order detail, at least for what concerns the SPARC sample.}

\subsection{Gas and stellar masses}

The final step to build the baryonic CoG is to choose appropriate mass-to-light ratios for the stellar disk ($\Upsilon_{\star,\,\rm disk}$) and the stellar bulge ($\Upsilon_{\star,\,\rm bul}$), as well as the factor to correct the \hi\ mass for the contribution of heavier elements ($M_{\rm gas}/M_{\hi}$). In analogy to previous SPARC papers, we assumed $\Upsilon_{\star,\,\rm disk}=0.5$, $\Upsilon_{\star,\,\rm bul}=0.7$, and $M_{\rm gas}/M_{\hi}=1.33$ for all galaxies. The choice of constant values provides the most empirical and data-driven representation of the data because it merely puts gas mass and stellar mass on a common physical scale, without any model-dependent variation from galaxy to galaxy (or within a given galaxy). Variations in $\Upsilon_{\star}$ are taken into account in our error budget, which considers uncertainties of 25$\%$ due to plausible differences in the galaxy star formation history and chemical enrichment history \citep[e.g.][]{Schombert2019-SFH, Schombert2022}.

It is possible to choose variable mass-to-light ratios \citep[e.g.][]{Schombert2022, Duey2025} and/or gas correction factors \citep{McGaugh2020} from galaxy to galaxy; we will investigate such potential improvements in future work.
As a preliminary study, we considered the values of $\Upsilon_{\rm \star}$ from fitting the spectral energy distribution (SED) of 110 SPARC galaxies \citep{Marasco2025}. We find that $\Upsilon_{\rm \star}$ from SED fitting have negligible effects on our final results because they deviate from our fiducial values only in gas-dominated dwarf galaxies (see Fig. 1 of \citealp{Marasco2025}), in which the stellar component plays little role in the values of $M_{\rm bar}$ and $R_{\rm 50, bar}$. This is in line with the fact that the scatter in the (stellar or baryonic) Tully-Fisher relation does not decrease using $\Upsilon_{\rm \star}$ from SED fitting, but rather increases with respect to the simple choice of a constant $\Upsilon_{\star}$ at 3.6 $\mu$m \citep{Ponomareva2018-BTFR, Marasco2025}.

Formally, our measurements of $M_{\rm bar}$ and $R_{\rm 50, bar}$ neglect molecular gas. On average, the molecular gas mass ($M_{\rm mol}$) of nearby galaxies is about 7$\%$ of the stellar mass \textcolor{black}{\citep{McGaugh2020, Saintonge2022}, so its contribution to $M_{\rm bar}$ is generally small (but the scatter in the $M_{\rm mol}/M_{\star}$ ratio can be significant, see \citealt{Calette2018}). Broadly speaking, molecular gas has a similar spatial distribution as the stellar disk on large (kpc) scales \citep{Leroy2008, Frank2016}, so its contribution could be implicitly included in the stellar density profile. On average, this} would be equivalent to assuming $\Upsilon_{\star,\,\rm disk}=0.535$ rather than $\Upsilon_{\star,\rm disk}=0.5$. \textcolor{black}{This small statistical correction would make no difference to our results because} it is smaller than expected variations in $\Upsilon_{\star}$, which are ultimately the most relevant uncertainties in the measurements of $M_{\rm bar}$ and $R_{\rm 50, bar}$.

Fig.~\ref{fig_derivation} shows the baryonic CoG for three characteristic galaxies: a gas-dominated case, a star-dominated case, and an intermediate case. For the gas-dominated galaxy, the baryonic half-mass radius is nearly identical to the gaseous half-mass radius ($R_{\rm 50, gas}$). Conversely, for the star-dominated galaxy, one has $R_{\rm 50, bar}\simeq R_{\rm 50, \star}$. For the intermediate case, $R_{\rm 50, bar}$ is somewhat in between $R_{\rm 50, gas}$ and $R_{\rm 50, \star}$, highlighting the importance of considering both stars and gas when studying galaxy sizes. \textcolor{black}{Notably, when the gas contribution is substantial, the baryonic CoG may not show a clear flattening as in the case of the stellar CoG because gas disks are generally more diffuse than stellar disks (see Fig.\,\ref{fig_derivation}). The \hi\ data used here reach \hi\ column densities of about $5-10 \times 10^{19}$ cm$^{-2}$ (the typical \hi\ sensitivity of historic radio interferometers), so they may not trace the full extent of the \hi\ disks.} \textcolor{black}{Recent, ultra-deep \hi\ observations with the MeerKAT telescope \citep[the MHONGOOSE survey;][]{deBlok2024, Marasco2025} show that the sizes of \hi\ disks do not increase substantially when probing column densities that are 2 orders of magnitude lower (a few times 10$^{17}$ cm$^{-2}$, see  Fig. 7 in \citealt{deBlok2024}), so some of our $M_{\rm bar}$ and $R_{\rm 50, bar}$ may be underestimated by $10\%-20\%$ at most, which is comparable to or smaller than our assumed uncertainties (see Appendix\,\ref{App_error}).}

\begin{figure*}
    \centering
    \includegraphics[width=1\linewidth]{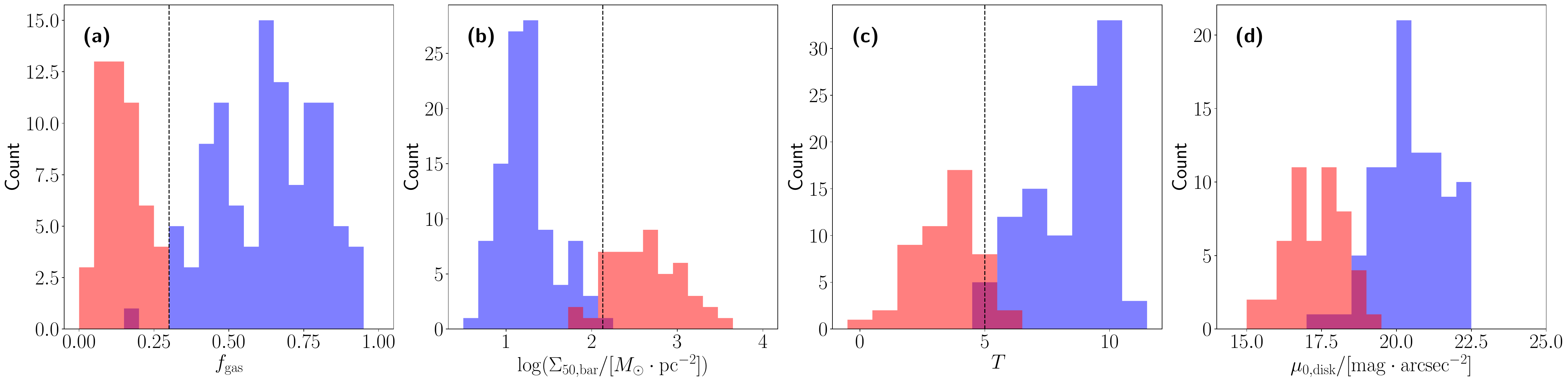}
    \caption{Distributions of $f_{\rm gas}$, $\Sigma_{\rm 50, bar}$, $T$, and $\mu_{\rm 0, disk}$ for the two groups identified by \textsc{DBSCAN} (red and blue histograms).
    The dashed black lines show the separation values reported in Table~\ref{tab_roxy_baryons}. Panels (a)-(c) show the distributions clustered by $\log\Sigma_{\rm 50, bar}$, while panel (d) shows the distributions clustered by $\mu_{\rm 0, disk}$ (after correction to face-on view).}
    \label{fig_cluster_hist}
\end{figure*}

\section{Results}\label{sec_MR}

\subsection{Two sequences in the baryonic mass-size plane}\label{sec:dbsan}

Figure~\ref{fig_gas-star-bar} shows the gaseous, stellar, and baryonic mass-size relations of SPARC galaxies, colour-coded by their gas fraction $f_{\rm gas} = M_{\rm gas}/M_{\rm bar}$. The gaseous mass-size relation (left panel) is a tight power law (in logarithmic space), in agreement with previous studies that used the `isophotal' \hi\ radius \citep[e.g.][]{Broeils1997, Verheijen2001, Swaters2002, Lelli2016-SPARC, Wang2016-M_HI-R_HI}. The stellar mass-size relation (middle panel) displays two distinct sequences, which were noted by \citet{Lelli2016-SPARC} in the luminosity-size plane (their Fig. 2) and are in agreement with the seminal work of \citet{Schombert2006-structure}. These two sequences approximately correspond to high surface brightness (HSB), star-dominated, early-type spirals (Sa to Sc) and to low surface brightness (LSB), gas-dominated, late-type disks (Sd to dI). The baryonic mass-size relation (right panel) makes the two sequences even more evident. The LSB sequence, indeed, becomes tighter because it mostly consists of gas-dominated objects, highlighting the importance of adding the gas component in the mass-size relation. If we use the radius that encompasses 20$\%$ or 80$\%$ of the baryonic mass ($R_{\rm 20, bar}$ and $R_{\rm 80, bar}$) instead of $R_{\rm 50, bar}$, two distinct HSB-LSB sequences persist, indicating that their existence does not depend on the specific definition of the baryonic radius. In the following, we consider baryonic surface densities rather than surface brightnesses, so we refer to the two distinct populations as high-surface-density (HSD) and low-surface-density (LSD) galaxies.

The existence of two distinct sequences is confirmed by the \textsc{DBSCAN} algorithm (\citealp{Ester1996-DBSCAN}), which can be used to identify clusters of points. This algorithm is robust against irregular boundaries and does not need a priori information about the number of clusters. We adopted the \textsc{DBSCAN} package in \textsc{Scikit-learn} \citep{Pedregosa2011-scikit}, and used it in a normalized five-dimensional space, consisting of $\log(M_{\rm bar})$, $\log (R_{\rm 50, bar})$, gas fraction $f_{\rm gas}$, Hubble type $T$, and effective surface density $\log(\Sigma_{\rm 50, bar})$. The gas fraction is defined as $f_{\rm gas}=M_{\rm gas}/M_{\rm bar}$, while the baryonic effective surface density as $\Sigma_{\rm 50, bar} = M_{\rm bar}/(2 \pi R_{\rm 50, bar}^2)$. Following \citet{Ester1996-DBSCAN}, we determined the best input parameters of \textsc{DBSCAN} as \texttt{Eps = 0.239} and \texttt{Min\_samples = 10}. \textcolor{black}{These two parameters work as follows: For a given data point in the parameter space, any other point within a radius equal to \texttt{Eps} is considered its neighbour. If the number of neighbours of a given point is greater than \texttt{Minpts}, the point will be considered as a `core point', so as the core of a cluster of points.}

As expected, \textsc{DBSCAN} divides our sample of galaxies into two main groups (plus a few outliers). Figure~\ref{fig_cluster_hist} shows that the two groups separate around $f_{\rm gas} \simeq 0.3$, $T \simeq 5$ (corresponding to Sc types), and $\log(\Sigma_{\rm 50, bar}/{\rm [M_\odot \, pc^{-2}]}) \simeq 2.1$.
\textcolor{black}{This clustering result is robust for a broad range of input parameters, specifically \texttt{Eps} $\in [0.18, 0.25]$ and \texttt{Minpts} $\in [7, 13]$.}
We then used the characteristic value $\Sigma_{\rm c}\simeq 125$ M$_\odot \, $pc$^{-2}$ to divide our sample into LSD ($\Sigma_{\rm 50, bar} < \Sigma_{\rm c}$) and HSD ($\Sigma_{\rm 50, bar} \geq \Sigma_{\rm c}$) galaxies. Using a characteristic $f_{\rm gas}$ or $T$ instead of $\Sigma_{\rm c}$ provides similar results in terms of separating two galaxy populations. Of course, any separation is a simplification because there is no sharp transition in galaxy properties: there are galaxies in-between the two sequences that are mostly Sc/Sd types with intermediate surface brightness and $f_{\rm gas}\simeq0.3-0.5$.

\begin{figure*}
    \centering
    \includegraphics[width=1\linewidth]{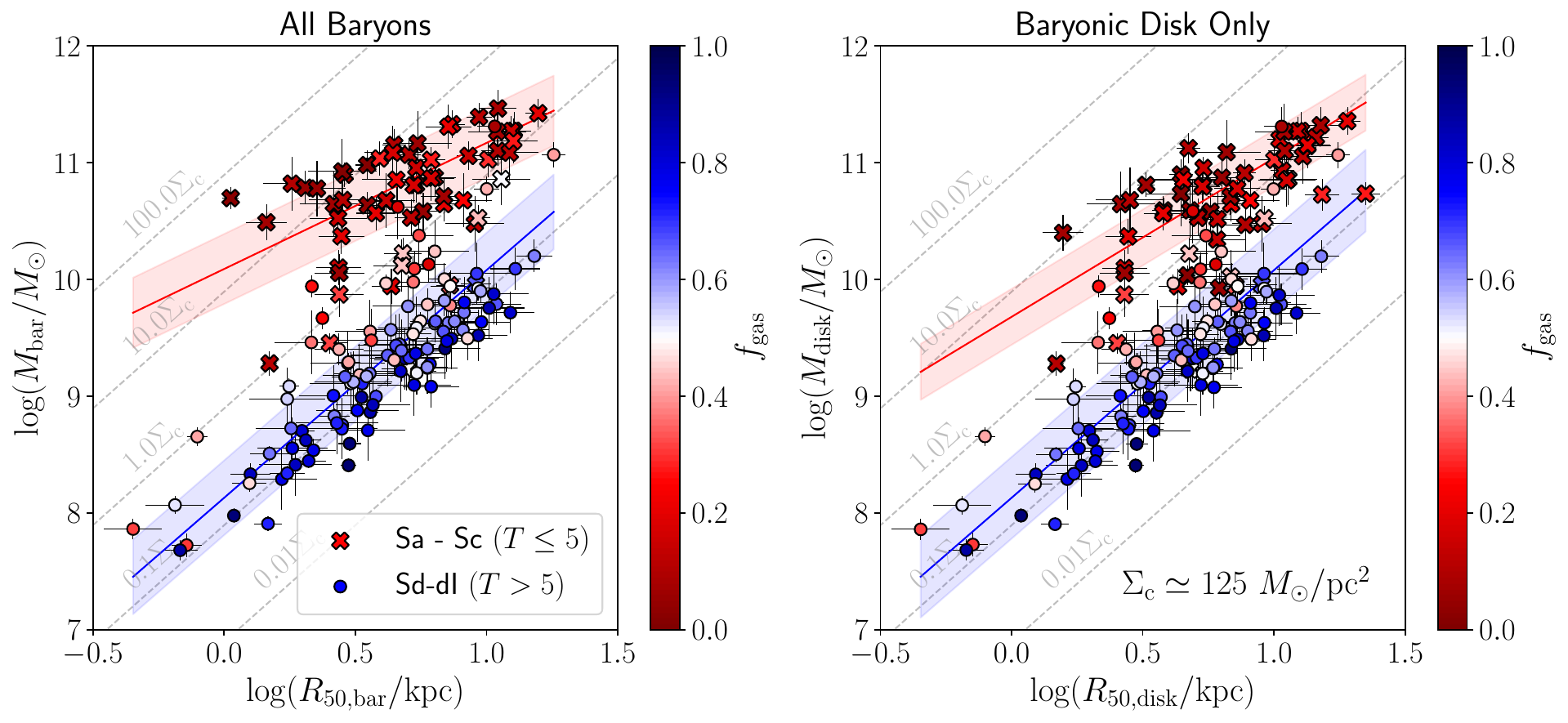}
    \caption{Baryonic mass-size relations considering all baryons (left panel) and the baryonic disks only (excluding stellar bulges, right panel). Galaxies are colour-coded by the gas fraction $f_{\rm gas}$. Crosses represent S0-to-Sc galaxies (Hubble type $T \leq 5$), while circles represent Sd-to-dI galaxies ($T > 5$). The red and the blue solid lines show the best-fit relations to the HSD and LSD sequences, respectively; the red and blue shaded areas indicate the best-fit intrinsic scatters. Dashed lines correspond to constant surface densities for multiples of $\Sigma_{\rm c}=125$ M$_\odot$ pc$^{-2}$ (see Sect. \protect{\ref{sec_MR}} for details).}
    \label{fig_main}
\end{figure*}

\subsection{Best-fit baryonic mass-size relations}
\label{sec_bestfit}
We fitted each galaxy group with a straight line,
\begin{equation}
    \log M_{\rm bar} = \alpha \log R_{\rm 50, bar} + \beta,
\end{equation}
where $\alpha$ is the slope, $\beta$ is the intercept. Fitting a straight line to data points with intrinsic scatter and errors on both variables is not a trivial exercise. For example, the \textsc{BayesLineFit} software \citep{Lelli2019-BTFR} uses a Markov-Chain Monte-Carlo approach in a Bayesian framework, assuming that the two variables are independent. In the case of the mass-size relation, however, the two variables are not independent because they both vary with galaxy distance $D$ ($M_{\rm bar} \propto D^2$ and $R_{\rm 50, bar} \propto D$). Thus, we used the \textsc{roxy} package (\citealp{Bartlett2023-roxy}) that can take the covariance between the variables into account. Details on the estimation of the covariance matrix are given in Appendix~\ref{App_error}. The current implementation of $\textsc{roxy}$, however, can fit only for the vertical intrinsic scatter, not the orthogonal intrinsic scatter perpendicular to the best-fit line. Appendix A of \citet{Lelli2019-BTFR} discusses the systematic differences between the two approaches (see also Appendix~\ref{App_ODR}).

The best-fit relations are shown in Fig.\,\ref{fig_main} (left panel), and the best-fit parameters are shown in Table~\ref{tab_roxy_baryons}. The baryonic mass-size relation of LSD galaxies has $\alpha = 1.95 \pm 0.12$, while that of HSD galaxies has $\alpha = 1.08 \pm 0.18$. The former value implies that LSD galaxies share a similar $\Sigma_{\rm 50, bar} \simeq 13$ M$_\odot \, $pc$^{-2}$, which is largely driven by the gas component. On the other hand, HSD galaxies become less dense and compact as their $M_{\rm bar}$ increase. Interestingly, the intrinsic scatter ($\sigma_{\rm int}$) of the HSD sequence ($\sigma_{\rm int} = 0.30 \pm 0.03$) and the LSD sequence ($\sigma_{\rm int}=0.32 \pm 0.02$) are consistent within the errors. The HSD sequence shows higher observed scatter (evident by eye) because $M_{\rm bar}$ is dominated by the stellar mass, which has higher uncertainties than the gas mass dominating $M_{\rm bar}$ in the LSD sequence. Intrinsically, however, the two sequences may be similarly tight.

\subsection{The effect of stellar bulges}\label{sec_bulges}

One may wonder whether the HSD-LSD dichotomy is driven by the presence of stellar bulges because HSD galaxies mostly have early Hubble types (Sa-Sc), whereas LSD galaxies are virtually bulgeless (Sd-dI). Indeed, \citet{Lelli2016-SPARC} performed bulge-disk decompositions for only two peculiar LSD galaxies out of 118, whereas bulge-disk decompositions were performed for 28 HSD galaxies out of 51 (mostly Sa-Sb types). We used these bulge-disk decompositions to measure the baryonic CoG of the baryonic disk (stars plus gas) and its half-mass radius, $R_{\rm 50, disk}$, excluding the bulge component. Importantly, the non-parametric bulge-disk decompositions of \citet{Lelli2016-SPARC} assign stellar bars, ovals, and small pseudo-bulges to the disk component. These structures are mostly present in HSD galaxies and are known to arise from the disk secular evolution, so they probably share a similar $\Upsilon_\star$ as the rest of the disk.

Fig.\,\ref{fig_main} (right panel) shows the baryonic-disk mass-size relation, excluding stellar bulges. The two HSD-LSD sequences are still evident by eye, and formally confirmed by the \textsc{DBSCAN} algorithm. Thus, the HSD-LSD dichotomy is not driven by the presence of stellar bulges. For LSD galaxies, the best-fit $M_{\rm disk}-R_{\rm 50, disk}$ relation is virtually the same as the $M_{\rm bar}-R_{\rm 50, bar}$ relation, as expected given the lack of stellar bulges. For HSD galaxies, instead, the best-fit $M_{\rm disk}-R_{\rm 50, disk}$ relation has a slightly steeper slope ($\alpha=1.36\pm0.16$) than the $M_{\rm bar}-R_{\rm 50, bar}$ relation ($\alpha=1.08\pm0.18$), but the two slopes are consistent within $1.16 \sigma$ (see also Appendix\,\ref{App_data}). This fact indicates that the details of the bulge-disk decompositions do not play an important role in our general results. Finally, we note that Sc/Sd galaxies (with small or no bulges) mostly fill the gap between the two sequences, suggesting that the onset of bulges could depend on the mean surface density of the baryonic disk. In other words, the existence of bulges in HSD disks may be a direct effect of the dichotomy, as we discuss in Sect \ref{sec_SB}.

\section{Discussion}\label{sec_diss}

\subsection{A dichotomy in star-forming galaxy disks}\label{sec_SB}

The existence of two sequences in the $M_{\rm bar}-R_{\rm 50, bar}$ plane suggests that there is a dichotomy in star-forming disks: galaxies tend to be either HSD or LSD, so tend to avoid intermediate surface densities around 100 M$_\odot$ pc$^{-2}$. It is then natural to ask whether the observed HSD-LSD dichotomy may be driven by potential selection effects. The SPARC sample is the result of decades of interferometric \hi\ observations from different groups \citep[see][for references]{Lelli2016-SPARC}, so it does not have a well-defined selection function. However, selection effects should generally bias against LSD or LSB galaxies, which are instead very well represented in SPARC. We cannot identify any sensible observational effect that would bias against intermediate surface densities. The upcoming BIG-SPARC database \citep{Haubner2024}, which will combine \hi\ observations with NIR photometry for thousands of galaxies, will clarify the situation.

Reassuringly, the same dichotomy was found by \citet{Schombert2006-structure} studying the stellar mass-size plane of a different galaxy sample. In addition, a similar dichotomy was found in terms of the disk central surface brightness $\mu_{\rm 0, disk}$ in other independent studies \citep{Tully1997-bimodality, McDonald2009-Bimodality_Ursa, McDonald2009-bimodality, Sorce2013-bimodality, Sorce2016-bimodality}. Clearly, a dichotomy in $\mu_{\rm 0, disk}$ has a similar physical meaning of a dichotomy in $\Sigma_{\rm 50, bar}$. Indeed, if we run \textsc{DBSCAN} replacing $\Sigma_{\rm 50, bar}$ with $\mu_{\rm 0, disk}$ (see Sect.\,\ref{sec:dbsan}), we still find two distinct groups of galaxies in the $M_{\rm bar}-R_{\rm 50, bar}$ plane (see Fig.\,\ref{fig_cluster_hist}, rightmost panel). The surface density of baryons (stars plus gas) is a more physical quantity than the surface brightness in some optical or NIR band. Moreover, it allows us to account for the dominant gas component in LSD galaxies, so we urge (when possible) to use $\Sigma_{\rm 50, bar}$ and the baryonic mass-size plane.

The dichotomy between HSD and LSD galaxies relates to other fundamental properties of galaxies, such as their gas fractions and gas depletion times. HSD galaxies tend to have low gas fractions ($\lesssim30\%$ of their baryonic mass), so have been very efficient in converting gas into stars during the Hubble time. Indeed, HSD galaxies will run out of gas in a few gigayears \citep[e.g.][]{McGaugh2017}, unless their gas reservoir is constantly replenished to sustain the current star formation rates \citep[e.g.][]{Sancisi2008}. On the other hand, LSD galaxies have high gas fractions ($30-90\%$ of their baryonic mass), so have been very inefficient in forming stars. Their gas depletion times are remarkably large. At their current star formation rates, LSD galaxies could keep forming stars for another 10-100 Gyr without any need of accreting new gas \citep[e.g.][]{vanZee2001, McGaugh2017}.

Another key difference between HSD and LSD galaxies regards their morphological properties, as indicated by their Hubble types. HSD galaxies (Sa-Sc) can have bulges, spiral arms, bars, and bar-driven structures (such as stellar lenses, rings, and pseudo-bulges). On the contrary, LSD galaxies usually lack these morphological features and their optical morphology is characterized by irregular and clumpy star formation. These morphological differences are likely driven by the different dynamical state of their baryonic disks. In HSD galaxies, the stellar disk is basically self-gravitating in the inner regions (near maximal, e.g. \citealt{vanAlbada1986, Starkman2018}), so it can sustain spiral arms and form stellar bars \citep{Lin1964-spiral, Sellwood2022-sipral_review}. In LSD galaxies, instead, the stellar disk is not self-gravitating: the baryonic mass is dominated by the gas, while the total mass is dominated by dark matter (in the standard cosmological paradigm) down to small radii \citep[e.g.][]{DeBlok1997, Tully1997-bimodality}. Both effects make the baryonic disk relatively stable against the propagation of density waves and formation of bars. Interestingly, the gap between the two populations may suggest that the co-domination of baryons and dark matter is avoided \citep{Tully1997-bimodality}.

The LSD-HSD dichotomy suggests that the two galaxy populations form and evolve along different paths. If the initial baryonic disk is relatively heavy and dense (the HSD case), the conversion of atomic gas in molecular gas may be facilitated, so the star formation process may be quite efficient. In addition, large-scale instabilities driven by the disk self-gravity may lead to the formation of spiral arms, bars, bulges, and pseudo-bulges \citep[e.g.][]{Lin1964-spiral, Sellwood2014, Sellwood2022-sipral_review}. These structures will then trigger gas condensations and shocks, thus efficient star formation and gas consumption (e.g. \citealp{Roberts1969-spiral, Kim2014-spiral, Yu2021-Spiral}). Conversely, if the initial baryonic disk is relatively light and diffuse (the LSD case), gravitational instabilities are less effective, resulting in inefficient star formation and abundant gas content \citep[e.g.][]{Wyder2009-SFR_LSBG}. Importantly, a LSD galaxy evolving in isolation will continue to be a LSD galaxy for at least another 10-100 Gyr. In this evolutionary scenario, the ability (or not) of sustaining density waves may be crucial because spiral arms are a very efficient mechanism to convert gas into stars \citep{Roberts1969-spiral, Yu2021-Spiral, Querejeta2024-Spiral}. It remains unclear, however, why there should be a gap between the two populations.

The gap between HSD and LSD may potentially represent a rare and rapid transitional stage, in which galaxies transform from LSD to HSD. Given the low star formation rates and high gas depletion times of LSD galaxies \citep[e.g.][]{McGaugh2017}, it will take more than a Hubble time for them to reduce their gas fractions towards those of HSD galaxies. The transformation of LSD into HSD galaxies (if any) must necessarily involve external mechanisms, such as galaxy interactions, which can trigger spiral arms, bars, and gas inflows \citep[e.g.][]{RamonFox-spiral_satellites, Lokas2019-spirals_Illustris}. Some studies \citep{Tully1997-bimodality, Sorce2013-bimodality} found that galaxies in the transition region tend to have close neighbours, suggesting that galaxy interactions may indeed play a key role in such a transformation.

\begin{figure*}
    \centering
    \includegraphics[width=1\linewidth]{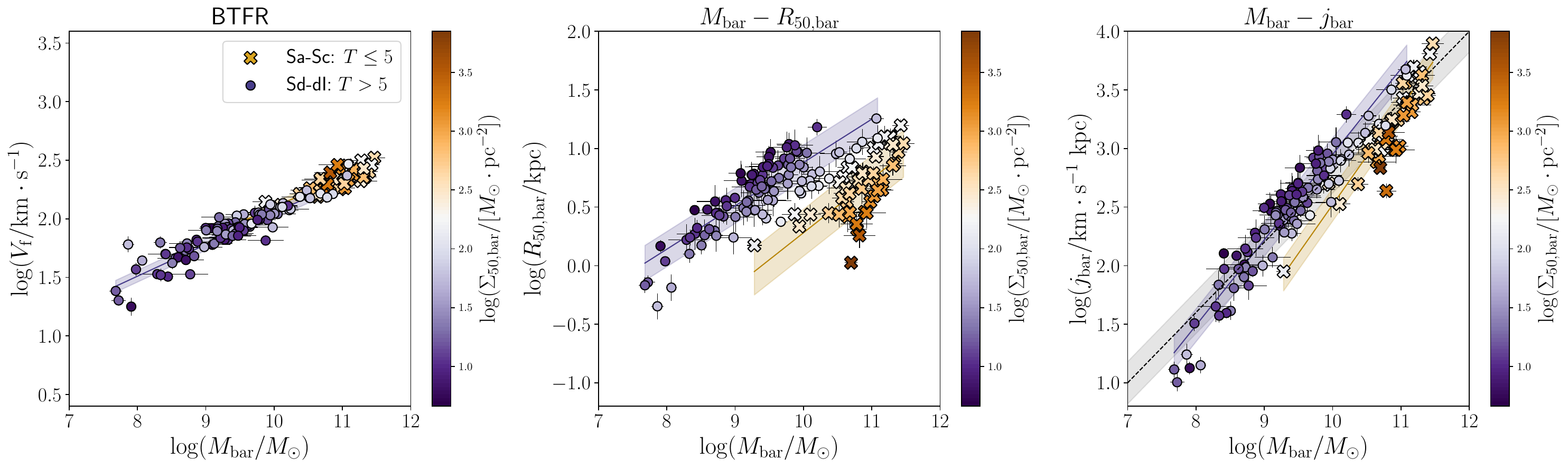}
    \caption{\textcolor{black}{Scaling relations of galaxy disks: the baryonic Tully-Fisher relation (left panel), the baryonic mass-size relation (middle panel), and the baryonic angular-momentum relation (right panel). The three panels cover the same dynamic range on the y-axis. The circles and squares are the same as those in Fig.~\protect{\ref{fig_main}}, colour-coded by the effective baryonic surface density $\Sigma_{\rm 50, bar}$. The blue and gold solid lines are the best-fit lines for LSD galaxies and HSD galaxies, respectively. The dashed black line in the right panel shows the best-fit line considering all 147 galaxies together.}}
    \label{fig_BTFR_MR_MJ}
\end{figure*}

\subsection{Relations with dynamical scaling laws}\label{sec_MJ}

One may wonder whether the HSD-LSD dichotomy is imprinted in other scaling relations of galaxies. In the following, we consider (1) the baryonic Tully-Fisher relation (BTFR) that links $M_{\rm bar}$ with the mean velocity along the flat part of the rotation curve $V_{\rm f}$ \citep[e.g.][]{Mcgaugh2000-BTFR, McGaugh2005-BTFR, Lelli2016-BTFR, Lelli2019-BTFR}; and (2) the angular momentum relation (AMR) that links $M_{\rm bar}$ with the specific angular momentum $j_{\rm bar}$ \citep[e.g.][]{ManceraPina2021-MJ1, ManceraPina2021-MJ}. To this aim, we adopted a subsample of \textcolor{black}{147} galaxies with high-quality rotation curves ($Q \neq 3$, see \citealp{Lelli2016-SPARC}) \textcolor{black}{and disk inclinations $i \geq 30^\circ$}. We took $V_{\rm f}$ from \citet{Lelli2019-BTFR}. For 27 galaxies without measured $V_{\rm f}$ (due to rising rotation curves up to the last measured point), we used the outermost velocity point as a first-order approximation, so that we have the largest possible sample to compare to the baryonic mass-size relation. Similarly, we considered the outermost value of $j_{\rm bar}(R)$ even if it has not converged to a constant value. Excluding these non-converging galaxies (39 out of 147) would not strongly affect the overall results of this section.

\textcolor{black}{For an axisymmetric disk galaxy with circular orbits, the specific angular momentum is defined as
\begin{equation}\label{eq_j}
    j_{\rm bar} = \dfrac{\int_0^{\infty} V(R) \cdot R \cdot \Sigma_{\rm bar}(R) \cdot 2 \pi R dR}{ \int_0^{\infty} \Sigma_{\rm bar}(R) \cdot 2 \pi R dR},
\end{equation}
where the denominator is equivalent to the total baryonic mass $M_{\rm bar}$. For a galaxy with a flat rotation curve, Eq.\,\ref{eq_j} reduces to
\begin{equation}\label{eq_j_Vf}
    j_{\rm bar} = \dfrac{V_{\rm f}}{M_{\rm bar}} \int_0^{\infty} \Sigma_{\rm bar}(R) \cdot 2 \pi R^2 dR = V_{\rm f} \cdot \mathcal{R}(M_{\rm bar}, R_{\rm 50, bar}),
\end{equation}
where $\mathcal{R}$ is a number with physical units of length (the result of the integral divided by $M_{\rm bar}$). In many practical circumstances, the baryonic mass distribution can be expressed as some function of $M_{\rm bar}$ and $R_{\rm 50, bar}$ (e.g. the S\'ersic profile), so $\mathcal{R}$ depends on these variables. For example, for an exponential profile for which the scale length $R_{\rm d}= R_{\rm 50, bar}/1.678$, Eq.\,\ref{eq_j_Vf} simplifies to the following \citep[c.f. with][]{Fall1983, Romanowsky2012}
\begin{equation}\label{eq_j_exp}
    j_{\rm bar} = 1.2 \cdot V_{\rm f} \cdot R_{\rm 50, bar}.
\end{equation}
In this approximation, the AMR is the mathematical product of the BTFR and the $M_{\rm bar}-R_{\rm 50, bar}$ relation, so it cannot contain extra information \citep[see the discussion in][]{Lelli2019-BTFR}. In the following, differently from \citet{Lelli2019-BTFR}, we computed $j_{\rm bar}$ in a more rigorous way by integrating Eq.\,\ref{eq_j} for gas and stars separately, following similar procedures as \citet{Posti2018-MJ} and \citet{ManceraPina2021-MJ1}. However, given the overall flatness of rotation curves at large radii, $j_{\rm bar}$ can be thought as $V_{\rm f} \cdot \mathcal{R}$ at an effective level to a first-order approximation.}

Figure~\ref{fig_BTFR_MR_MJ} shows the three scaling relations. Importantly, the vertical axes cover the same dynamic range in $V_{\rm f}$, $R_{\rm 50, bar}$, and $j_{\rm bar}$, so that the scatter around the relations and any eventual dichotomy can be visually assessed. It is evident that the tightest among the three is the BTFR, as already pointed out and quantified in \citet{Lelli2019-BTFR}.

The BTFR does not distinguish between LSD and HSD galaxies. Even if we fit the two galaxy populations separately, we recover the same best-fit relation within the uncertainties. This is in agreement with previous studies, which do not find additional correlations between the BTFR residuals and other galaxy properties \citep[e.g.][]{Lelli2016-BTFR, Lelli2019-BTFR, Ponomareva2018-BTFR, Desmond2019, Hua2024-BTFR}. \textcolor{black}{In particular, claims of LSB galaxies deviating from the BTFR \citep{ManceraPina2019-UDG_BTFR, Hu2023-UDG_BTFR, Du2024-UDG, Rong2024-UDG} are very dubious because they are driven by face-on galaxies with low-quality data and/or by inappropriately comparing \hi\ line-widths from spatially unresolved data with well-measured $V_{\rm flat}$ from rotation curves \citep[see][]{Lelli2024-UDG}. Empirically, the BTFR seems to have irreducible scatter, so that considering any extra variable in addition to $V_{\rm f}$ at fixed $M_{\rm bar}$ (such as Eq.\,\ref{eq_j_Vf}) would only degrade the original correlation.}

The AMR displays a moderate dichotomy: HSD and LSD galaxies follow different AMRs with a small offset in normalization. This offset must be driven by differences in the baryonic mass distribution at fixed $M_{\rm bar}$ \textcolor{black}{because the BTFR shows no residual correlations with size and/or surface density (cf. with Eq.\,\ref{eq_j_Vf}). This occurs not only at an `asymptotic' level using $V_{\rm flat}$, but also at a `local' level using the velocity $V(R)$ at each radius $R$:} the various $M_{\rm bar}-V(R)$ relations display no residual correlations with size at fixed mass \citep{Desmond2019}. In other words, the dichotomy in $M_{\rm bar}-j_{\rm bar}$ plane is the same as the one in the $M_{\rm bar}-R_{\rm 50, bar}$ plane, but appears less evident because \textcolor{black}{the optimal variable ($V_{\rm f}$) that gives a near-perfect correlation with $M_{\rm bar}$ is effectively multiplied by another variable (such as $\mathcal{R}$ in Eq.\,\ref{eq_j_Vf} or $R_{\rm 50, bar}$ in Eq.\,\ref{eq_j_exp}).}

The moderate dichotomy in the AMR is in agreement with the results of \citet{ManceraPina2021-MJ}, who find that gas-rich galaxies have a larger intercept than gas-poor ones. Gas fraction is known to correlate with surface brightness or surface density (see, e.g. Fig,\,\ref{fig_gas-star-bar}), so the offset found by \citet{ManceraPina2021-MJ} is the same as the HSD-LSD dichotomy. Importantly, the $j_{\rm bar}-M_{\rm bar}-f_{\rm gas}$ plane proposed by \citet{ManceraPina2021-MJ} may be the result of a circular argument. Essentially, one takes an intrinsically thin relation (the BTFR) and convolves it with another variable (such as $\mathcal{R}$ or $R_{\rm 50, bar}$ at an effective level) to obtain a new relation with increased scatter (the AMR). Next, the increased scatter in the AMR is reduced considering a fourth variable ($f_{\rm gas}$) that correlates with surface density and/or $R_{\rm 50, bar}$ at fixed $M_{\rm bar}$, effectively removing the extra dependency that was introduced in building the AMR in the first place.

The previous reasoning is confirmed by a simple mathematical exercise. The `effective' AMR relation is given by
\begin{equation}
    \log(V_{\rm f} \cdot R_{\rm 50, bar}) = S \cdot \log(M_{\rm bar}) + N
\end{equation}
where the slope $S$ is the sum of the slopes of the BTFR and the  $M_{\rm bar}-R_{\rm 50, bar}$ relation. By fitting LSD and HSD galaxies separately, we find that $S \sim$0.64 for LSD galaxies and $\sim $0.72 for HSD galaxies. These values are very similar to the best-fit slopes of the actual AMR relation ($\sim$0.73 for LSD and $\sim$0.83 for HSD galaxies). Of course, the values are not exactly the same because $j_{\rm bar}$ is not calculated as $V_{\rm f} \times R_{\rm 50, bar}$ but by using Eq.\,\ref{eq_j}. Nevertheless, this simple exercise highlights the fact that $j_{\rm bar}$ is largely set by $V_{\rm f}$ and $R_{\rm 50, bar}$ in typical galaxies. In other words, only two scaling relations among the BTFR, AMR, and $M_{\rm bar}-R_{\rm 50, bar}$ are independent. The BTFR is the tightest one and shows no correlation with other galaxy properties ($R_{\rm 50, bar}$, $\Sigma_{\rm 50, bar}$, and so on), so it is appropriate to consider the BTFR as a primary relation.

In the context of galaxy formation and evolution, it is sensible to think that $j_{\rm bar}$ determines $V_{\rm f}$ and $R_{\rm 50, bar}$ \citep[e.g.][]{Mo1998-disk, Fall1980-formation}. However, taking the BTFR as a primary relation, the AMR and the $M_{\rm bar}-R_{\rm 50, bar}$ relation must contain nearly the same empirical information. Then, from a practical perspective, the $M_{\rm bar}-R_{\rm 50, bar}$ plane appears to be a better probe of galaxy evolution than $M_{\rm bar}-j_{\rm bar}$ because the two galaxy populations (HSD and LSD galaxies) are much better separated in the former one, leaving little space for ambiguity. In conclusion, among the three scaling relations, the combination of BTFR and $M_{\rm bar}-R_{\rm 50, bar}$ seems to be the best choice to test models of galaxy formation and evolution. In particular, the challenge for $\Lambda$CDM models is to reproduce the tightness of the BTFR while having two distinct sequences in the $M_{\rm bar}-R_{\rm 50, bar}$ plane of disk galaxies.

\subsection{Connections with Milgromian dynamics}
\label{sec_MOND}
The previous discussion is largely independent of the current cosmological model and the nature of dark matter. Our findings, however, have a potential connection with Milgromian dynamics (or Modified Newtonian dynamics, MOND), a paradigm that modifies the standard laws of gravity and/or inertia rather than adding dark matter \citep{Milgrom1983-MOND}. MOND effects occur below an acceleration scale $a_0\simeq1.2 \times 10^{-10}$ m s$^{-2}$, which is imprinted in the dynamical scaling laws of galaxies \citep[e.g.][]{Lelli2017-RAR, Lelli2022}. The acceleration scale $a_0$ can be recast as a characteristic surface density scale $\Sigma_{\rm M}= a_0/(2\pi G)\simeq 137 \ M_{\rm \odot}/{\rm pc^2}$ \citep[e.g.][]{Milgrom2016, Milgrom2024}, where $G$ is Newton's constant. 

The central density relation (CDR) of galaxies is particularly relevant in this context \citep{Lelli-2016-CDR}. The CDR links the central `observed' baryonic surface density $\Sigma_{\rm 0, bar}$ with the `dynamical' surface density $\Sigma_{\rm 0, dyn}$ inferred from the inner steepness of the rotation curve. The CDR is a non-linear relation that displays a knee around $\Sigma_{\rm M}$. Galaxies with $\Sigma_{\rm 0, bar} > \Sigma_{\rm M}$ are along the one-to-one relation, so they are baryon-dominated in the inner parts (in the Newtonian regime in a MOND context). Galaxies with $\Sigma_{\rm 0, bar} < \Sigma_{\rm M}$ systematically deviate from the one-to-one line, so they are progressively more and more DM-dominated in the inner parts (in the Milgromian regime in a MOND context).

Interestingly, the division between HSD and LSD galaxies ($\Sigma_{\rm c}\simeq 125$ M$_{\odot}$/pc$^2$) from the baryonic mass-size plane is practically indistinguishable from the characteristic MOND surface density $\Sigma_{\rm M} \simeq 137 \ M_{\rm \odot}/{\rm pc^2}$. Approximately, using the adequate mass-to-light conversions, this corresponds to the historical Freeman's limit for the central surface brightness of galaxy disks \citep{Freeman1970-Disk}. This fact is quite remarkable because MOND is a theory of dynamics; yet the presence of $a_0$ is also imprinted on the structural properties of galaxies, such as the baryonic mass-size relation and the baryonic surface density distribution, which do not involve any kinematic measurement. 

In the MOND context, the HSD-LSD dichotomy may naturally arise because the theory has a characteristic surface density scale, $\Sigma_{\rm M}$, which distinguishes between different physical regimes. Galaxies (or proto-galaxies) with $\Sigma_{\rm 50, bar}\gtrsim \Sigma_{\rm M}$ are in the Newtonian regime in the inner regions (with no dark matter), so they are prone to bar-like instabilities \citep{Ostriker1973}. By internal disk evolution, they will increase their central surface densities until they possibly reach a near-stable configuration due to the formation of bars and pseudo-bulges \citep{Combes2014, Nagesh2023}. Thus, galaxy disks with $\Sigma_{\rm 50, bar}\simeq \Sigma_{\rm M}$ would necessarily be rare because they have evolved towards higher $\Sigma_{\rm 50, bar}$, forming the HSD sequence. On the other hand, galaxy disks with $\Sigma_{\rm 50, bar} < \Sigma_{\rm M}$ are in the deep MOND regime, which increases their stability. These galaxies will form the LSD sequence and remain there for most of their lifetime, consuming gas in a slow, inefficient fashion (unless some external mechanisms alter their internal stability). Thus, LSD galaxies are expected to have low SFRs and low metallicities, as observed \citep[e.g.][]{Bothun1997}. For LSD galaxies, the mean value of $\Sigma_{\rm 50, bar}\simeq 10-15$ M$_\odot$ pc$^{-2}$ must then be set by the hydrodynamical properties of atomic gas, rather than by stellar dynamics. Large sets of hydrodynamical simulations of galaxy formation in MOND \citep[e.g.][]{Combes2014, Wittenburg2020, Nagesh2023} are necessary to further test these ideas.

This evolutionary scenario may be closely related to the initial baryonic angular momentum in MOND. As pointed out by \citet{Milgrom2021-MOND_J}, MOND defines a fiducial specific angular momentum that depends on the baryonic mass of the galaxy:
\begin{equation}
j_{\rm M}(M_{\rm bar}) = \left(\frac{G^3 M_{\rm bar}^3}{a_0}\right)^{1/4} \simeq 382 \left(\dfrac{M_{\rm bar}}{10^{10} M_\odot} \right)^{3/4} \rm{km\,s^{-1}\,kpc},
\end{equation}
A proto-galaxy with mass $M_{\rm bar}$ and $j_{\rm bar}\gg j_{\rm M}$ will settle into a LSD disk. Instead, a proto-galaxy with mass $M_{\rm bar}$ but $j_{\rm bar}\lesssim j_{\rm M}$ will settle into a HSD disk with $M_{\rm disk} = (j_{\rm bar}/j_{\rm M}) M_{\rm bar}$ and develop a low-$j$ component (such as a pseudo-bulge) that takes up the rest of the mass. In addition, \citet{Milgrom2021-MOND_J} shows that
\begin{equation}\label{eq_Sigma_j}
    j_{\rm bar} \simeq \sqrt{\frac{\Sigma_{\rm disk}}{\Sigma_{\rm M}}} j_{\rm M} \simeq 382 \left(\frac{\Sigma_{\rm disk}}{\Sigma_{\rm M}} \right)^{1/2} \left(\dfrac{M_{\rm bar}}{10^{10} M_\odot} \right)^{3/4} \rm{km\,s^{-1}\,kpc},
\end{equation}
where $\Sigma_{\rm disk}$ is the mean surface density of the baryonic disks, which can be approximated by $\Sigma_{\rm 50, disk}$ within a factor of $O(1)$.

Our results are in overall agreement with the MOND predictions. First, HSD disks are systematically below the AMR relation defined by LSD disks (Fig.\,\ref{fig_BTFR_MR_MJ}, right panel). Second, gas-dominated LSD disks in our sample have $\Sigma_{\rm 50, bar}\simeq 0.1 \Sigma_{\rm M}$ (Fig.\,\ref{fig_main}), so MOND predicts that their AMR relation should have a slope of $3/4=0.75$, which is entirely consistent with the best-fit slope of \textcolor{black}{$0.72\pm0.02$}. For star-dominated HSD disks, instead, the slope of the AMR relation ($0.83 \pm 0.06$) seems to deviate from the MOND prediction, but the situation is more difficult to quantitatively assess. There are two main reasons: (1) HSD disks are mostly made of stars, so the computation of $j_{\rm bar}$ is a very rough approximation because we do not have access to the actual stellar kinematics (through spectroscopic observations), but we use \hi\ rotation curves to infer the stellar rotation and stellar angular momentum assuming an average pressure term in the Jeans equations \citep[similarly to][]{Posti2018-MJ}; (2) HSD disks must undergo a redistribution of their angular momentum through secular evolution due to the formation of bars and pseudo-bulges, so the modelling of these components in the computation of $j_{\rm bar}$ and $\Sigma_{\rm 50, disk}$ becomes critical; for example, our bulge-disk decompositions assign bars and pseudo-bulges to the disk (because they are expected to have a similar stellar mass-to-light ratios) so they are included in $j_{\rm bar}$ and $\Sigma_{\rm 50, disk}$, which is different from the definitions given by \citet{Milgrom2021-MOND_J}.

\section{Conclusion}\label{sec_conclusion}

We introduced the baryonic mass-size relation of SPARC galaxies, which links the total baryonic mass (stars plus gas) with the baryonic half-mass radius. We find the following results:

\begin{enumerate}
    \item SPARC galaxies follow two distinct sequences in the $M_{\rm bar}-R_{\rm 50, bar}$ plane: one defined by HSD, star-dominated, Sa-Sc galaxies and one defined by LSD, gas-dominated, Sd-dI galaxies. The two sequences, which are evident by eye, are confirmed by the clustering algorithm \textsc{DBSCAN}.
    \item The baryonic mass-size relation of LSD galaxies has a slope close to 2, pointing to a constant average baryonic surface density of the order of 10-15 $M_\odot$ pc$^{-2}$. The baryonic mass-size relation of HSD galaxies has a slope close to 1, indicating that less massive spirals are progressively more compact. The same results hold if we consider baryonic disk-only relations, excluding stellar bulges from the computation.
    \item The HSD-LSD dichotomy is totally absent in the baryonic Tully-Fisher relation ($M_{\rm bar}$ versus $V_{\rm flat}$) but moderately seen in the angular-momentum relation ($M_{\rm bar}$ versus $V_{\rm flat}\times R_{\rm 50, bar}$), so it is driven by variations in the baryonic distribution ($R_{\rm 50, bar}$) at fixed baryonic mass. This fact indicates that these relations are not independent, and that the baryonic mass-size relation provides a more incisive probe of evolutionary differences rather than the angular-momentum relation.
\end{enumerate}

Our results confirm the early findings of \citet{Schombert2006-structure} for a different galaxy sample. In addition, the HSD-LSD dichotomy is in line with previous studies that found a dichotomy in terms of the disk central surface brightness \citep{Tully1997-bimodality, Sorce2013-bimodality, Sorce2016-bimodality}. To put these results on firmer grounds, we are currently building a much larger sample of galaxies with both spatially resolved \hi\ data and NIR photometry \citep[BIG-SPARC,][]{Haubner2024}. 

In a companion paper (Hua et al. in preparation), we show how studying the baryonic mass-size relation in conjunction with the stellar mass-size relation provides key insights in possible morphological transformations between star-forming and passive galaxies. In particular, we add passive galaxies (ellipticals, lenticulars, dwarf ellipticals, and dwarf spheroidals) to the current sample to obtain a global view on galaxy evolution.

\begin{acknowledgements}
    The authors thank Konstantin Haubner and Illaria Ruffa for constructive discussions during this study. EDT was supported by the European Research Council (ERC) under grant agreement no. 101040751.
\end{acknowledgements}

\bibliographystyle{aa}
\bibliography{MRrelation}

\begin{appendix}
\section{Tables}\label{App_data}
In section~\ref{sec_method}, we described the compilation of \hi\ surface density profiles for 169 galaxies. References for the \hi\ data and the corresponding number of galaxies are listed in Table~\ref{tab_references}.

In section~\ref{sec_MR}, we used the \textsc{roxy} software to fit the baryonic mass-size relations of HSD and LSD galaxies separately. Table~\ref{tab_roxy_baryons} provides the best-fit parameters of the $M_{\rm bar}-R_{\rm 50, bar}$ relation, considering different physical properties to divide the two groups of galaxies. Table~\ref{tab_roxy_disk} provides the same information for the $M_{\rm disk}-R_{\rm 50, disk}$ relation, in which stellar bulges are excluded.

In Section~\ref{sec_diss}, we used the \textsc{roxy} software to fit the baryonic Tully-Fisher relation, the baryonic mass-size relation (switching the axes with respect to Sect.\,\ref{sec_MR}), and the baryonic angular momentum relation for HSD and LSD galaxies separately. Table~\ref{tab_MJ} provides the best-fit parameters.

\begin{table}[h]
\centering
\caption{References for the \hi\ surface density profiles}
\label{tab_references}
\begin{tabular}{l r}
  \hline
  \hline
  Source & Number of galaxies\\
  \hline
  \citet{Swaters2002} & 37 \\
  \citet{Verheijen2001} & 27 \\
  \citet{deBlok1996} & 16 \\
  \citet{Noordermeer2005} & 12 \\
  \citet{Richards2015, Richards2016} & 9 \\
  \citet{Lelli2014a} & 8 \\
  \citet{Broeils1992} & 7 \\
  \citet{Begeman1987} & 6 \\
  \citet{vanderhulst1993} & 6 \\
  \citet{vanZee1997} & 6 \\
  \citet{Spekkens2006} & 5 \\
  \citet{Gentile2004} & 4 \\
  \citet{Cote2000} & 3 \\
  \citet{Begum2003, Begum2005} & 2 \\
  \citet{Fraternali2011} & 2 \\
  \citet{Hallenbeck2014} & 2 \\
  \citet{Carignan1990_N7793, Carignan1990_N247} & 2 \\
  \citet{Begum2004} & 1 \\
  \citet{Barbieri2005} & 1 \\
  \citet{Boomsma2008} & 1 \\
  \citet{Carignan1988} & 1 \\
  \citet{Carignan1989} & 1 \\
  \citet{Chemin2006} & 1 \\
  \citet{Cote1991} & 1 \\
  \citet{Jobin1990} & 1 \\
  \citet{Kepley2007} & 1 \\
  \citet{Lake1990} & 1 \\
  \citet{Puche1991} & 1 \\
  \citet{Rhee1996} & 1 \\
  \citet{Roelfsema1985} & 1 \\
  \citet{VerdesMontenegro1997} & 1 \\
  \citet{Walsh1997} & 1 \\
  \hline
\end{tabular}
\end{table}

\begin{table*}[h]
    \begin{center}
    \caption{\textsc{roxy} best-fit parameters for the $R_{\rm 50, bar}-M_{\rm bar}$ relations, dividing SPARC galaxies in two groups based on $\Sigma_{\rm 50, bar}$ (our fiducial distinction between HSD and LSD galaxies), $f_{\rm gas}$, and $T$.}
    \label{tab_roxy_baryons}
    \begin{tabular}{c |c c| c c| c c}
        \hline
        \hline
         & $\Sigma_{\rm 50, bar} \geq \Sigma_{\rm M}$&$\Sigma_{\rm 50, bar} < \Sigma_{\rm M}$ &$f_{\rm gas} \leq 0.3$&$f_{\rm gas} > 0.3$ & $T \leq 5$ & $T > 5$\\
         \hline
        $\alpha$ & $1.08 \pm 0.18$& $1.95 \pm 0.12$  & $1.03 \pm 0.14$ & $1.86 \pm 0.12$ & $0.85 \pm 0.23$ & $1.83 \pm 0.13$ \\
        $\beta$ & $10.09 \pm 0.13$ & $8.13 \pm 0.08$ & $10.07 \pm 0.13$ & \textcolor{black}{$8.16 \pm 0.08$} & $10.11 \pm 0.07$ & $8.19 \pm 0.08$\\
        $\sigma_{\rm int}$ & $0.30 \pm 0.03$ & $0.32 \pm 0.02$ & $0.35 \pm 0.05$ & \textcolor{black}{$0.34 \pm 0.02$} & $0.45 \pm 0.03$ & $0.27 \pm 0.02$\\
        \hline
    \end{tabular}
    \end{center}
\end{table*}

\begin{table*}[h]
    \begin{center}
    \caption{\textsc{roxy} best-fit parameters for the $R_{\rm 50, disk}-M_{\rm disk}$ relations, dividing SPARC galaxies in two groups based on \textcolor{black}{$\Sigma_{\rm 50, bar}$ (our fiducial distinction between HSD and LSD galaxies), $f_{\rm gas}$, and $T$.}}
    \label{tab_roxy_disk}
    \begin{tabular}{ c |c c| c c| c c }
        \hline
        \hline
         & $\Sigma_{\rm 50, bar} \geq \Sigma_{\rm M}$&$\Sigma_{\rm 50, bar} < \Sigma_{\rm M}$ &$f_{\rm gas} \leq 0.3$&$f_{\rm gas} > 0.3$ & $T \leq 5$ & $T > 5$\\
         \hline
        $\alpha$ & $1.36 \pm 0.16$ & $1.95 \pm 0.12$ & $1.33 \pm 0.17$ & $1.87 \pm 0.12$ & $1.27 \pm 0.23$ & $1.85 \pm 0.12$\\
        $\beta$ & $9.68 \pm 0.13$ & $8.13 \pm 0.08$ & $9.66 \pm 0.13$ & $8.16 \pm 0.08$ & $9.70 \pm 0.19$ & $8.18 \pm 0.08$\\
        $\sigma_{\rm int}$ & $0.24 \pm 0.04$ & $0.35 \pm 0.02$ & $0.28 \pm 0.03$ & $0.33 \pm 0.02$ & $0.39 \pm 0.04$ & $0.31 \pm 0.02$\\
        \hline
    \end{tabular}
    \end{center}
\end{table*}

\begin{table*}[h]
    \begin{center}
    \caption{\textsc{roxy} best-fit parameters of the baryonic Tully-Fisher relation, baryonic mass-size relation, and baryonic angular-momentum relation fitting LSD galaxies (left columns) and HSD galaxies (right columns) separately.}
    \label{tab_MJ}
    \begin{tabular}{ c |c c| c c| c c}
        \hline
        \hline
        & \multicolumn{2}{c|}{BTFR: $M_{\rm bar}$ vs $V_{\rm f}$ } & \multicolumn{2}{c|}{Mass-size: $M_{\rm bar}$ vs $R_{\rm 50, bar}$} & \multicolumn{2}{c}{AMR: $M_{\rm bar}$ vs $j_{\rm bar}$}\\
        & HSD & LSD & HSD & LSD & HSD & LSD\\
        \hline
        $\alpha$ & $0.26 \pm 0.02$& $0.27 \pm 0.01$ 
                 & $0.46 \pm 0.08$ & $0.36 \pm 0.02$ 
                 & $0.83 \pm 0.06$ & $0.72 \pm 0.02$\\
        $\beta$ & $-0.51 \pm 0.22$ & $-0.65 \pm 0.10$ 
                & $-4.34 \pm 0.82$ & $-2.74 \pm 0.21$ 
                & $-5.78 \pm 0.68$ & $-4.29 \pm 0.22$ \\
        $\sigma_{\rm int}$ & $0.04 \pm 0.01$& $0.05 \pm 0.01$ 
                           & $0.19 \pm 0.02$ & $0.15 \pm 0.01$ 
                           & $0.14 \pm 0.02$ & $0.14 \pm 0.01$\\
        \hline
    \end{tabular}
    \end{center}
\end{table*}

\section{Error estimation}\label{App_error}

Following \citet{Lelli2016-BTFR}, we estimated the error on $M_{\rm bar}$ as
\begin{equation}\label{eq_err-Mbar}
 \delta_{M_{\rm bar}} = \sqrt{\left( \dfrac{\delta_{F_{\rm HI}}}{F_{\rm HI}} M_{\rm gas}\right)^{2} + \left(\dfrac{\delta_{L}}{L} M_\star\right)^{2} + \left( \dfrac{\delta_{\Upsilon_{*}}}{\Upsilon_{\star}} M_\star\right)^{2} + \left(2 \dfrac{\delta_{D}}{D}M_{\rm bar}\right)^{2}},
\end{equation}
where $\delta_{F_{\rm HI}}$ is the uncertainty on the \hi\, flux (about 10$\%$), $\delta_{L}$ is the uncertainty on the 3.6 $\mu$m luminosity (a few per cent), $\delta_{\Upsilon_{\rm *}}$ is the uncertainty on the stellar mass-to-light ratio (dominated by galaxy-to-galaxy variations that are assumed to be of $\sim$25$\%$), and $\delta_{D}$ is the uncertainty on the galaxy distance (varying from 5$\%$ to 30$\%$ depending on the galaxy, see \citealt{Lelli2016-SPARC}). \textcolor{black}{When a galaxy bulge is present, the third term inside the square root was replaced by two similar terms: one for the stellar disk and one for the stellar bulge.} The error on the disk mass (excluding the bulge) was given by the same equation by replacing $M_\star$ with the mass of the stellar disk and $M_{\rm bar}$ with the mass of the baryonic disk.

The errors on $R_{\rm 50, bar}$ were estimated as
\begin{equation}\label{eq_eR}
    \delta_{R_{\rm 50, bar}} = \sqrt { \left(\frac{\delta_{\theta_{50}}}{\theta_{50}} R_{\rm 50, bar} \right)^2+ \left(\frac{\delta_D}{D} R_{\rm 50, bar} \right)^2}, 
\end{equation}
where $\theta_{50}$ is half-mass radius in arcsec. We assumed that $\delta_{\theta_{50}}/\theta_{50} = 1/n$, where $n$ is the number of \hi\ spatially resolved elements along the disk semi-major axis (approximately the number of points along the SPARC rotation curve).

In logarithmic space, the covariance between $\delta_{M_{\rm bar}}$ and $\delta_{R_{\rm 50, bar}}$ is given by
\begin{equation}\label{eq_cov}
    \sigma_{M_{\rm bar}R_{\rm 50,bar}} = 2 \left(\frac{\delta_D}{D \ln 10}\right)^2.
\end{equation}

The values of $V_{\rm f}$ and the corresponding errors are taken directly from \citet{Lelli2019-BTFR}. For galaxies without a measured $V_{\rm f}$, we considered the last point of the rotation curve and the corresponding error (accounting for uncertainties on inclination too). $V_{\rm f}$ does not depend on $D$, so there is no need to consider the covariance matrix between $M_{\rm bar}$ and $V_{\rm f}$.

The error on $j_{\rm bar}$ is estimated following \citet{ManceraPina2021-MJ}. The covariance between $M_{\rm bar}$ and $j_{\rm bar}$ (in logarithmic space) can be roughly estimated as,
\begin{equation}
    \sigma_{M_{\rm bar}j_{\rm bar}} = \sigma_{M_{\rm bar}R_{\rm 50,bar}},
\end{equation}
according to equation~(\ref{eq_j_exp}). Nevertheless, we find that even if we do not include the covariant term $\sigma_{M_{\rm bar}j_{\rm bar}}$, the best-fitting results do not change significantly.

\section{Results using different fitting codes}
\label{App_ODR}

In Section~\ref{sec_bestfit}, we determined the best-fit parameters of the $R_{\rm 50, bar}-M_{\rm bar}$ relation using the \textsc{roxy} software, which fits for the vertical intrinsic scatter (`vertical fitting'). It is known that the results produced by vertical and orthogonal fittings can have considerable differences (see Appendix A of \citealp{Lelli2019-BTFR}), so we repeated the fits using the codes \textsc{BayesLineFit} \citep{Lelli2019-BTFR} and \textsc{HyperFit} \citep{Robotham2015-Hyperfit}, both of which allow for orthogonal fitting.

The results are provided in Tables~\ref{tab_Bayeslinefit_baryons} and \ref{tab_Hyperfit_baryons}, respectively. Both orthogonal fitting codes yield larger slopes than those given by the vertical fitting code \textsc{roxy}, in agreement with the results of \citet{Lelli2019-BTFR}. In general, it is not obvious to decide whether vertical fitting is preferable to orthogonal fitting (or vice versa). In this paper, we adopted the results from vertical fitting because \textsc{roxy} has been amply tested using mock datasets and seems to provide unbiased results \citep{Bartlett2023-roxy}.

Finally, we note that the best-fit parameters from \textsc{HyperFit} and \textsc{BayesLineFit} are consistent with each other within $1\sigma$, albeit the former takes the covariance matrix of each data point into account whereas the latter does not. This confirms that the covariance matrix has a minor effect on the best-fit results.

\begin{table*}[h]
    \begin{center}
    \caption{\textsc{BayesLineFit} best-fit parameters of the $R_{\rm 50, bar}-M_{\rm bar}$ relations, dividing SPARC galaxies in two groups based on $\Sigma_{\rm 50, bar}$ (our fiducial distinction between HSD and LSD galaxies), $f_{\rm gas}$, and $T$.}
    \label{tab_Bayeslinefit_baryons}
    \begin{tabular}{c |c c| c c| c c}
        \hline
        \hline
         & $\Sigma_{\rm 50, bar} \geq \Sigma_{\rm M}$&$\Sigma_{\rm 50, bar} < \Sigma_{\rm M}$ &$f_{\rm gas} \leq 0.3$&$f_{\rm gas} > 0.3$ & $T \leq 5$ & $T > 5$\\
         \hline
        $\alpha$
                 & $2.01^{+0.34}_{-0.27}$& $2.51 ^{+0.15}_{-0.14}$
                 & $2.32^{+0.46}_{-0.33}$ & $2.40^{+0.15}_{-0.14}$
                 & $3.88^{+1.29}_{-0.85}$ & $2.50^{+0.19}_{-0.16}$ \\
        $\beta$  & $9.46^{+0.19}_{-0.24}$ & $7.80^{+0.09}_{-0.10}$
        & $9.19^{+0.23}_{-0.33}$ & $7.84^{+0.09}_{-0.10}$
                & $7.98^{+0.60}_{-0.90}$ & $7.80^{+0.11}_{-0.12}$\\
        $\sigma_{\rm int,\perp}$ 
                           & $0.16^{+0.02}_{-0.02}$ & $0.13^{+0.01}_{-0.01}$ 
                           & $0.18^{+0.02}_{-0.02}$ & $0.12^{+0.01}_{-0.01}$ 
                           & $0.22^{+0.03}_{-0.02}$ & $0.14^{+0.01}_{-0.01}$\\
        \hline
    \end{tabular}
    \tablefoot{We used the orthogonal fitting, which provides the intrinsic scatter perpendicular to the best-fit relation.}
    \end{center}
\end{table*}

\begin{table*}[h]
    \begin{center}
    \caption{\textsc{HyperFit} best-fit parameters of the $R_{\rm 50, bar}-M_{\rm bar}$ relations, dividing SPARC galaxies in two groups based on $\Sigma_{\rm 50, bar}$ (our fiducial distinction between HSD and LSD galaxies), $f_{\rm gas}$, and $T$.}
    \label{tab_Hyperfit_baryons}
    \begin{tabular}{c |c c| c c| c c}
        \hline
        \hline
         & $\Sigma_{\rm 50, bar} \geq \Sigma_{\rm M}$&$\Sigma_{\rm 50, bar} < \Sigma_{\rm M}$ &$f_{\rm gas} \leq 0.3$&$f_{\rm gas} > 0.3$ & $T \leq 5$ & $T > 5$\\
         \hline
        $\alpha$  
                 & $1.82^{+0.26}_{-0.23}$& $2.51 ^{+0.15}_{-0.14}$  
                 & $2.06^{+0.37}_{-0.28}$ & $2.39^{+0.14}_{-0.13}$
                 & $3.09^{+0.78}_{-0.55}$ & $2.45^{+0.17}_{-0.15}$ \\
        $\beta$ & $9.60^{+0.16}_{-0.19}$ & $7.79^{+0.09}_{-0.10}$
        & $9.37^{+0.20}_{-0.26}$ & $7.84^{+0.09}_{-0.09}$
                & $8.54^{+0.40}_{-0.56}$ & $7.81^{+0.10}_{-0.11}$\\
        $\sigma_{\rm int,\perp}$ 
                           & $0.17^{+0.02}_{-0.02}$ & $0.14^{+0.01}_{-0.01}$ 
                           & $0.20^{+0.02}_{-0.02}$ & $0.14^{+0.01}_{-0.01}$ 
                           & $0.23^{+0.03}_{-0.02}$ & $0.16^{+0.01}_{-0.01}$\\
        \hline
    \end{tabular}
    \tablefoot{We used the orthogonal fitting, which provides the intrinsic scatter perpendicular to the best-fit relation.}
    \end{center}
\end{table*}

\end{appendix}

\end{document}